%% file: explain.tex
\documentclass[sigconf, nonacm, screen, pbalance]{acmart}
\usepackage{hyperref}
\usepackage[inline]{enumitem}
\usepackage{subcaption}
\usepackage{booktabs}
\usepackage{tabularx}
\usepackage{nicematrix}
\usepackage{cleveref}
\usepackage[normalem]{ulem}

\usepackage[newfloat=true,frozencache=true]{minted}
\newcommand{\sparagraph}[1]{\vspace{1mm}\noindent {\bf #1}}

\usepackage{overpic}
\usepackage{tikz}
\newcommand*\captionlabel[1]{%
  \tikz[baseline=(char.base)]{%
    \node[shape=circle,fill=white,draw=black,inner sep=1.5pt, outer sep=5pt] (char) {%
      \textcolor{black}{\scriptsize \textsf{#1}}};
}}

\definecolor{codebg}{RGB}{248,248,248}
\setminted[sql]{
  fontsize=\small,
  bgcolor=codebg,
  breaklines,
  autogobble,
  tabsize=2,
  frame=none,
}
\input{constants}

\begin{document}
\title{EXPLAIN Yourself! Finding Query Planner Stalls Across DBMSes}

\author{Geoffrey X. Yu}
\orcid{0009-0005-3186-1465}
\affiliation{%
  \institution{MIT CSAIL}
  \city{Cambridge}
  \state{MA}
  \country{USA}
}
\email{geoffxy@mit.edu}

\author{Ryan Marcus}
\orcid{0000-0002-1279-1124}
\affiliation{%
  \institution{University of Pennsylvania}
  \city{Philadelphia}
  \state{PA}
  \country{USA}
}
\email{rcmarcus@seas.upenn.edu}

\author{Tim Kraska}
\orcid{0009-0003-2414-2759}
\affiliation{%
  \institution{MIT CSAIL}
  \city{Cambridge}
  \state{MA}
  \country{USA}
}
\email{kraska@mit.edu}

\input{sections/0-abstract}

\maketitle

\input{sections/1-introduction}
\input{sections/2-overall}
\input{sections/3-common}
\input{sections/4-related}
\input{sections/5-conclusion}

\bibliographystyle{ACM-Reference-Format}
\bibliography{explain_full}

\end{document}

%% file: constants.tex
\newcommand{\dbmsA}{DBMS A}
\newcommand{\dbmsB}{DBMS B}
\newcommand{\dbmsC}{DBMS C}
\newcommand{\dbmsD}{DBMS D}

\newcommand{\duckdb}{DuckDB}
\newcommand{\mysql}{MySQL}
\newcommand{\postgres}{PostgreSQL}

%% file: sections/0-abstract.tex
\begin{abstract}
Query planners are typically expected to produce optimized plans quickly, leading
many researchers (including the authors of this paper) and practitioners to
design systems that assume query planning is a low-cost operation.
Using a lightweight agentic search, we show that this assumption does not always
hold. Across seven DBMSes, including four commercial systems, we find at least
one query per system that takes more than three minutes to plan.
In addition to being slow to plan, such queries risk tying up database resources
without performing useful work, creating a potential denial-of-service vector.
We analyze the queries our search uncovers and compare how the seven systems
respond to each pattern.
We find that although the queries triggering slow planning are largely
DBMS-specific, recurring pathologies involving correlated subqueries, CTE
expansion, repeated subquery expressions, disjunctive joins, and constant
folding affect multiple systems.
We release our uncovered queries along with a curated suite of parameterized
query pathologies that researchers and database engineers can use to test
planner robustness.
Overall, our results show that query planning cannot always be treated as a
predictably inexpensive operation and that its latency and robustness deserve
further attention from both database researchers and engineers.
\end{abstract}

%% file: sections/1-introduction.tex
\section{Introduction}\label{sec:introduction}
Query planners are designed to \emph{quickly} produce physical plans. Recently,
researchers (including the authors of this paper) and practitioners have treated
this design goal as an assumption, developing components that assume query plans
can be cheaply generated.
For example, learned cost models and query run time predictors, which are used
to make time-sensitive decisions such as query scheduling, featurize
physical query plans~\cite{t3-rieger25, stage-wu24, lce-sun19, qppnet-marcus19,
zeroshot-hilprecht22, unify-wu22, neo-marcus19, bao-marcus22} in order to make
their predictions.
This is not only an academic assumption; production-grade schedulers like
Auto-WLM~\cite{autowlm-saxena23} and query performance predictors like
Stage~\cite{stage-wu24} both rely on featurizing query plans and making fast
decisions under single-digit milliseconds~\cite{stage-wu24}.

\begin{listing}[t]
\centering
\begin{minted}{sql}
SELECT COUNT(*) FROM lineitem AS l1
  WHERE EXISTS (SELECT 1 FROM lineitem AS l2
    WHERE l2.l_orderkey = l1.l_orderkey
      AND l2.l_suppkey <> l1.l_suppkey)
  AND NOT EXISTS (SELECT 1 FROM lineitem AS l3
    WHERE l3.l_suppkey <> l1.l_suppkey
      AND l3.l_receiptdate > l3.l_commitdate
      AND l3.l_orderkey = l1.l_orderkey)
\end{minted}
\caption{This TPC-H-derived query takes about 14~\underline{\smash{minutes}} to
\texttt{EXPLAIN} on \dbmsA{}, but 6~milliseconds on \duckdb{}.
\vspace{-1.5em}
}
\label{lst:slow-example}
\end{listing}

Unfortunately, this assumption is unsafe in practice. Consider the query shown
in \Cref{lst:slow-example}.
Despite taking only 6 milliseconds to plan on DuckDB, this query takes
\emph{over 14 minutes} to plan on an unnamed~\cite{dewitt_clause} commercial
database system, whose vendor is aware of this behavior. This example shows how
even a conceptually simple query can take surprisingly long to plan in
practice.

We call such instances of slow query planning a \emph{query planner stall}.
In addition to delaying query execution, stalls raise system robustness concerns
because the planner may consume substantial CPU time or memory before even
executing the query. Furthermore, a malicious user could expend planner
resources simply by issuing adversarial \texttt{EXPLAIN} commands, creating a
potential denial-of-service risk: a risk amplified by agentic workloads which
are expected to issue arbitrary and novel queries.  Planner stalls could thus
threaten robustness even when no user intentionally constructs an adversarial
query.

In this work, we demonstrate that stalls occur across diverse DBMSes and can be
easily generated: we give an LLM agent a single tool that measures the planning
time of a submitted SQL query and ask it to find slow-planning queries.
Despite the simplicity of this approach, the agent uncovers planner stalls on
all seven DBMSes we tested, including four commercial systems.
\textbf{On all seven systems, we find at least one query that takes more than
three minutes to plan.}
Some of the queries we find are not obviously contrived or adversarial; indeed,
we encountered one such query incidentally while investigating an unrelated
problem~\cite{tailwind-yu26}.
Our results suggest that query planner stalls are not just an isolated problem
in a single DBMS, but rather a recurring phenomenon across many modern,
production-grade DBMSes.

We study the queries uncovered by our agent and identify recurring pathologies
involving correlated subqueries, CTEs, repeated subquery expressions,
disjunctive joins, and constant folding.
Across these cases, we find that planner transformations and search procedures
can cause planning time to grow sharply, and sometimes exponentially, as modest
amounts of query structure are added.
Which patterns trigger stalls varies across DBMSes, suggesting that these
queries are not necessarily fundamentally difficult to plan, but instead expose
problematic cases in a specific planner's optimization strategy or
implementation.

Looking broadly, our findings motivate future research on practical query
planner robustness. A robust query planner should guarantee that it produces a
usable plan within a bounded time, degrade gracefully when optimization becomes
difficult, and resist adversarial inputs that trigger pathological planning
behavior. Similarly, downstream systems that rely on physical plans should
tolerate delayed or unavailable plans.
Toward this goal, we contribute (i) an analysis of query pathologies that lead
to planner stalls on existing DBMSes, and (ii) our uncovered queries, along with
a suite of parameterized query pathology families distilled from our analysis,
for testing planner
robustness.\footnote{\url{https://github.com/geoffxy/explain-yourself/}}
We will also release the lightweight agentic search tool we used to find the
queries.

%% file: sections/2-overall.tex
\section{Finding Query Planner Stalls}\label{sec:stalls}
\begin{figure}[t]
  \begin{overpic}[width=0.95\columnwidth, clip, trim={0cm 0.4in 0cm 0cm}]{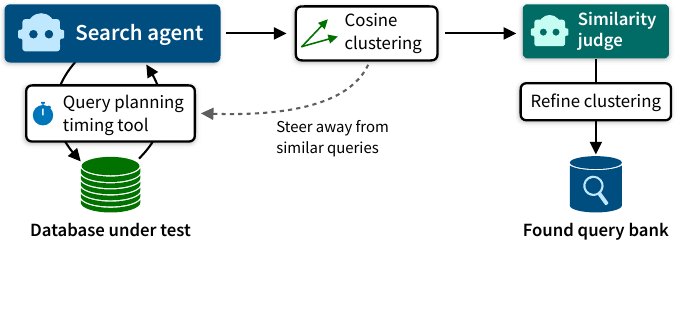}
    \put(14,26){\captionlabel{A}} 
    \put(35,35){\captionlabel{B}} 
    \put(35,16){\captionlabel{C}} 
    \put(70.5,21.5){\captionlabel{D}} 
  \end{overpic}
  \vspace{-1em}
  \caption{An overview of our agentic search procedure.}
  \label{fig:search-workflow}
  \vspace{-1em}
\end{figure}
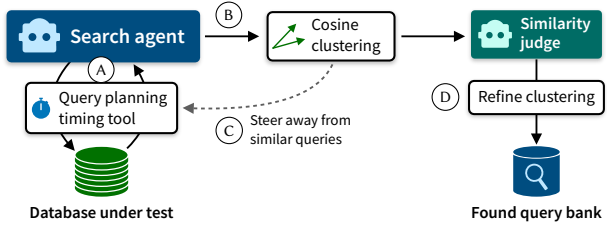

We start with an overview of our agentic search procedure
(\Cref{fig:search-workflow}) and then analyze the slow-to-plan queries we have
uncovered.

\subsection{Search Procedure}

We deliberately use a simple agentic search procedure to find query planner
stalls; our focus is on the resulting planner behaviors rather than the search
procedure itself.
We provide an LLM with a single tool that, given a SQL query, measures and
returns the query planning time on the database system being
tested~\captionlabel{A}.
We then instruct the LLM to use the tool to find and return as many queries as
it can whose planning time exceeds some threshold (e.g., 3 minutes).
If the LLM returns queries whose times do not exceed a minimum planning time
threshold (5 seconds in our experiments), we instruct it to try again.
Otherwise, we save the qualifying queries and start a new search attempt.
We run this search for $N$ turns (we use $N = 200$ in our experiments); one turn
corresponds to one LLM call.

\sparagraph{De-duplication.}
When the search agent finds queries~\captionlabel{B}, we de-duplicate them to
(i) avoid storing structurally similar queries, and (ii) to discourage
re-discovering known patterns. De-duplicating is nontrivial: exact or fuzzy SQL
equality may not match syntactically different queries that stress the same
planner behavior.

We represent each query using a feature vector containing its numbers of AST
nodes, joins, CTEs, window functions, aggregations, predicates, and set
operations. We then consider two queries similar when their vectors' cosine
similarity is at least $\delta$.
We use this threshold to gather queries into clusters and take the query with
the longest observed planning time from each cluster. We use cosine similarity
to cluster queries with similar structural shapes even when they contain
proportionally different numbers of operations.

We also use this similarity test during the search. Before timing a candidate
query $q$, the timing tool compares it with previously found
queries~\captionlabel{C}. If $q$ has a cosine similarity at least $\delta$ with
any archived query, the tool reports that it is too similar to a known query,
prompting the agent to explore a different candidate.

\sparagraph{Query bank.}
Cosine similarity produces an inexpensive first-pass clustering, but some
semantically similar queries may still end up in different clusters.
To refine this clustering, we use an LLM judge to assign a semantic similarity
score $S$ to every pair of remaining queries. We then
group queries such that every pair within a group has $S \ge
\tau$~\captionlabel{D}. We use $\tau=3$ on a five-point scale, where 5 indicates
that two queries are identical.
Finally, we manually remove queries expected to be slow to plan because they
only join excessively many tables or compute a \texttt{CUBE} over many columns.

\sparagraph{Database systems.}
We test seven database systems, four of which are commercial systems. We list
the DBMSes in \Cref{tbl:dbms}, anonymized when necessary~\cite{dewitt_clause},
with a description of their design and intended workload.
We run the commercial DBMSes in the cloud, each with 4 vCPUs and 32 GiB of memory.
We use the default configuration on each DBMS.
We obtain a query’s planning time on a DBMS by measuring how long the DBMS takes
to complete an \texttt{EXPLAIN} command (or its equivalent) for that query.

\begin{table}[t]
\small
\centering
\caption{The database systems tested in this work.}
\vspace{-1em}
\label{tbl:dbms}
\begingroup
\setlength{\tabcolsep}{4pt}
\begin{tabularx}{\columnwidth}{@{}llX@{}}
\toprule
\textbf{System} & \textbf{Commercial?} & \textbf{Design and intended workload} \\
\midrule
\dbmsA{} & Yes & Distributed analytical database \\
\dbmsB{} & Yes & Single-node, transactions and analytics \\
\dbmsC{} & Yes & Single-node, transactions and analytics \\
\dbmsD{} & Yes & Single-node, transactional \\
\midrule
\duckdb{} & No & Embedded in-process, in-memory analytics \\
\mysql{} & No & Single-node, transactional \\
\postgres{} & No & Single-node, transactional \\
\bottomrule
\end{tabularx}
\endgroup
\vspace{-1em}
\end{table}

\sparagraph{Dataset.}
We run our experiments on TPC-H data~\cite{tpch} using scale factor 100.
We use this dataset for its familiarity within the database community.
However, note that none of the pathologies described in this work are specific
to TPC-H.

\begin{figure}[t]
  \includegraphics[width=\columnwidth]{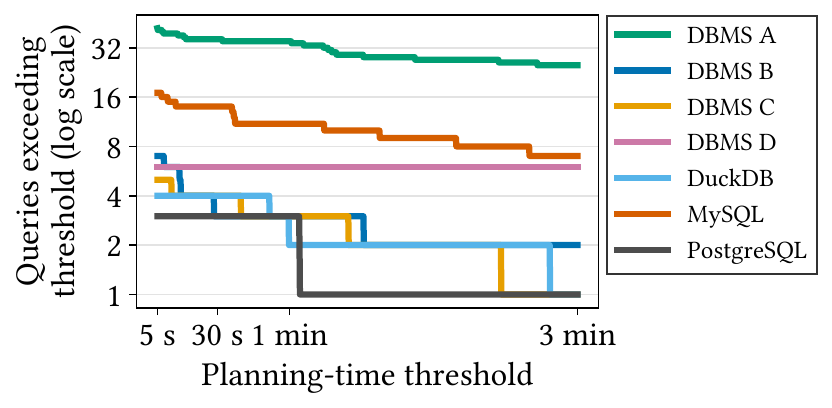}
  \vspace{-2.5em}
  \caption{Number of distinct query patterns found to take beyond increasing
    thresholds of time to plan.}
  \label{fig:search-overall}
  \vspace{-1em}
\end{figure}

\begin{figure*}[t]
  \includegraphics[width=0.95\textwidth]{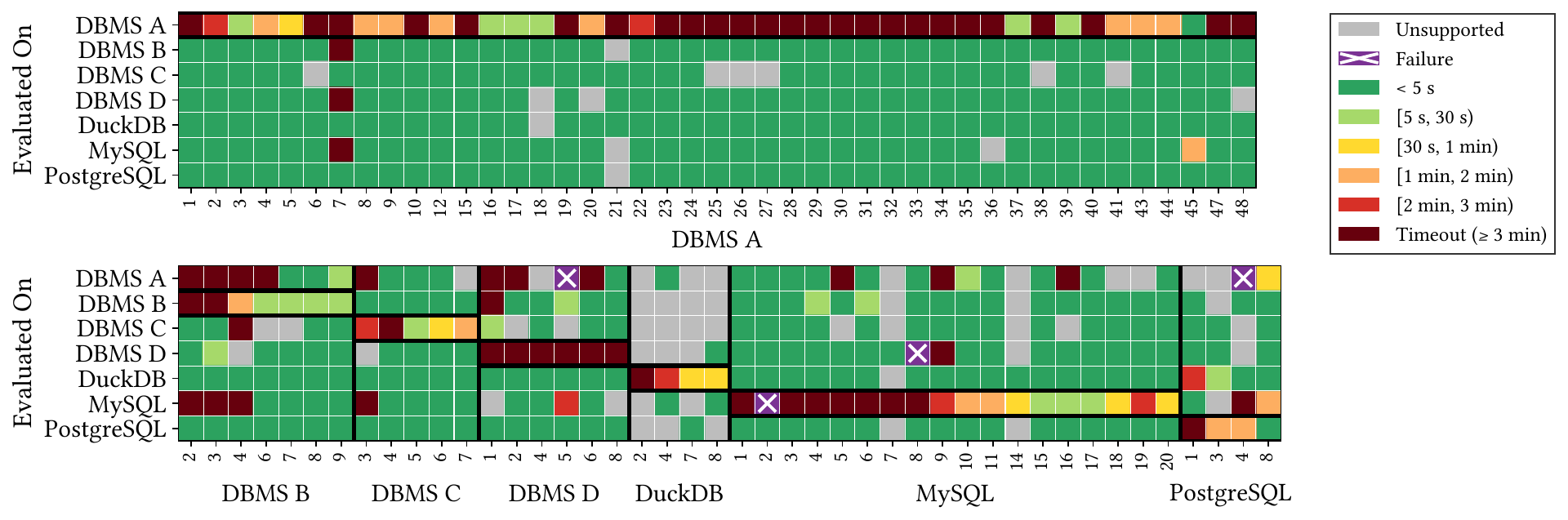}
  \vspace{-1.25em}
  \caption{A cross-system evaluation of the slow-planning queries our search
  tool found. Columns show queries grouped by the DBMS where they were
  discovered; rows show the DBMS used to plan them. The cell colors indicate
  planning latency or failure. A few queries transfer across DBMSes, but most
  seem to only trigger slow planning behavior on their discovery system.}
  \label{fig:cross-plot}
  \vspace{-0.75em}
\end{figure*}

\subsection{Search Results}
We use GPT-5.5 as our search agent, and GPT-5.4 Nano as our similarity judge to
reduce cost.
We run our search agent with a slow threshold of 3 minutes, 1 minute, and 30
seconds. \Cref{fig:search-overall} shows the number of queries we find
exceeding these planning time thresholds (in log scale). We make the following
observations.

\textbf{Query planner stalls occur on all seven DBMSes; stalls are not localized
problems specific to one database system.}
We find at least one query that took over 3 minutes to plan on each DBMS. At
the upper threshold of 3 minutes, \dbmsC{}, \duckdb{}, and \postgres{} admit
the fewest slow-planning queries (one query) whereas \dbmsA{} admits the most
(25 queries).

\textbf{The discovered query planner stalls vary in count and severity across
the DBMSes.}
\dbmsA{}, \dbmsD{}, and \mysql{} retain many slow-planning queries as we
increase the threshold from 5 seconds to 3 minutes, indicating a substantial
tail. In contrast, \dbmsC{}, \dbmsB{}, \duckdb{}, and \postgres{} exhibit a
similar number of queries with planning times above 5 seconds, but their counts
decline sharply as the slow threshold increases.

We note that these results reflect the query patterns that our search agent
discovered and they should not be interpreted as a ranking of the DBMSes'
quality. While differences across DBMSes could reflect their susceptibility to
planner stalls, they could also be due to biases in the queries explored by our
search agent.

\subsection{Do Planner Stalls Transfer Across Systems?}
Next, we see if a query found to trigger slow planning on one DBMS also triggers
a stall on other DBMSes.
We take each query that our search agent found on one DBMS and measure its
planning time on the six other DBMSes, translating SQL dialects as needed.
\Cref{fig:cross-plot} shows our results. The columns
represent each query found on each corresponding DBMS. The rows represent the
DBMS on which we measured the planning time; the diagonal, outlined in black,
represents queries planned on the same DBMS on which they were discovered. Gray
boxes indicate the query is unsupported on the DBMS; this usually means the
query uses a feature (e.g., a function) that the other DBMS does not support.

\textbf{Most slow-planning queries are specific to the DBMS on which they were
discovered.} 
This is most noticeable on \dbmsA{} and \mysql{}, as the darkest cells are
along the diagonal whereas most of the corresponding cells on the other DBMSes
are green.
In other words, queries that take minutes to plan on one system often plan in
under five seconds elsewhere. This finding suggests that stalls are likely due
to DBMS-specific planner behavior, rather than queries that are fundamentally
difficult to plan.

\textbf{Nevertheless, there do exist queries found on one DBMS that are also
slow to plan on others.}
For example, queries 2--4 on \dbmsB{} also lead to timeouts on \dbmsA{} and
\mysql{}. We see similar examples on the other six DBMSes.

%% file: sections/3-common.tex
\section{Query Planner Pathologies}
Although most slow-planning queries are specific to the DBMS on which they were
discovered, we do see a few pathological query patterns that affect more than
one DBMS. We look at five such pathologies in more detail next, starting with
correlated subqueries.

\begin{listing}[t]
\begin{minted}{sql}
SELECT COUNT(*) FROM lineitem AS l1 WHERE
  l1.l_receiptdate > l1.l_commitdate
  AND EXISTS (
    SELECT 1 FROM lineitem AS l2 WHERE
      l2.l_orderkey = l1.l_orderkey
      -- Alternates between <> and =
      AND l2.l_suppkey <> l1.l_suppkey
      -- Alternates between l_receiptdate and l_shipdate
      AND l2.l_receiptdate > l2.l_commitdate
  )
  AND EXISTS ( ... ) ...
\end{minted}
\caption{The correlated subquery pathology. We vary the number of
\texttt{EXISTS (...)} predicates from 0 to 15.}
\label{lst:corr-subq}
\end{listing}

\begin{figure}[t]
  \vspace{-1em}
  \includegraphics[width=\columnwidth]{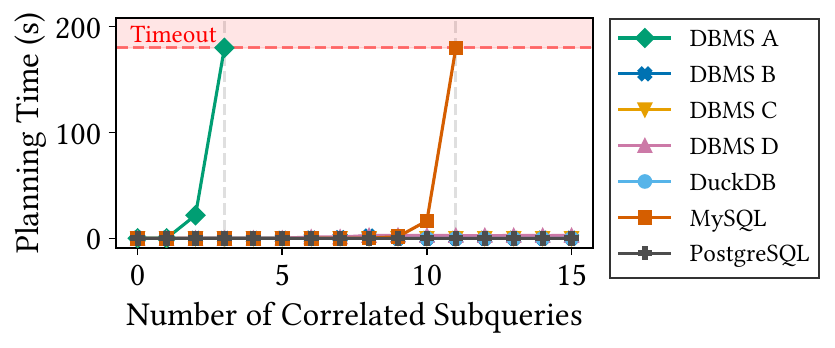}
  \vspace{-2.5em}
  \caption{\dbmsA{} and \mysql{} exhibit sharp increases in planning times as
  the number of correlated subqueries increases.}
  \label{fig:theme-correlated}
  \vspace{-1.5em}
\end{figure}

\subsection{Correlated Subqueries}
Correlated subqueries are subqueries that reference columns from an outer query
and planners generally try to decorrelate them~\cite{unnesting}.

\Cref{lst:corr-subq} shows a recurring pathology in our search, which comprises
correlated subqueries inside \texttt{EXISTS} filter predicates.
To study how this pathology affects planning performance, we plan the query on
all seven DBMSes while increasing the number of correlated subqueries from 1 to
15. We construct each query variant by adding copies of the \texttt{EXISTS}
predicate, alternating between the comparators and columns in the correlated and
non-correlated predicates respectively. \Cref{fig:theme-correlated} shows our
results using a three-minute planning timeout.
Both \dbmsA{} and \mysql{} are affected: \dbmsA{} reaches the timeout with
three correlated subqueries, while \mysql{} reaches it at 11.
The other five DBMSes appear to be unaffected.

\sparagraph{Why does this pathology occur?}
We investigated the planning slowdown by inspecting \mysql{}'s planner code
and its online documentation.
\mysql{} optimizes this query pattern by converting the \texttt{EXISTS}
subqueries into semi-joins and then enumerating join orders. By
default, \mysql{} sets its ``optimizer search depth'' to 62~\cite{mysql-depth}.
When this depth is greater than the number of tables in the query, it
exhaustively searches the join-order space~\cite{mysql-ordering-code}, which
contains $O(n!)$ orderings for $n$ tables.
If we manually set the optimizer search depth to 1 or disable \mysql{}'s
semi-join optimizations, this query takes single-digit milliseconds to plan. 
Thus \mysql{}'s sharp increase in planning time is because it performs
exhaustive join order enumeration on this query.
\dbmsA{} is closed-source, so we cannot inspect its code to investigate why
this pathology occurs.

\begin{listing}[t]
\begin{minted}{sql}
WITH v1 AS (SELECT r_regionkey AS rk FROM region
    WHERE r_regionkey = 0),
  v2 AS (SELECT i1.rk AS rk FROM v1 i1 JOIN v1 i2
    ON i1.rk = i2.rk),
  v3 AS (SELECT i1.rk AS rk FROM v2 i1 JOIN v2 i2
    ON i1.rk = i2.rk), ...
SELECT * FROM vn
\end{minted}
\vspace{-1em}
\caption{The CTE expansion pathology. Each CTE joins the previous CTE to itself
  and projects one column.}
\label{lst:cte-expansion}
\end{listing}

\subsection{Adversarial CTE Expansion}
A second pathology that affects several DBMSes involves repeatedly referenced
common table expressions (CTEs).
\Cref{lst:cte-expansion} shows a simplified query pattern that captures the core
problem. The \texttt{v1} CTE produces a single column and single row (there is
exactly one row with \texttt{r\_regionkey = 0}). Each subsequent CTE joins the
previous CTE with itself and projects the first column. Despite its syntactic
complexity, this query produces just a single value (0).

\begin{figure}[t]
  \includegraphics[width=\columnwidth]{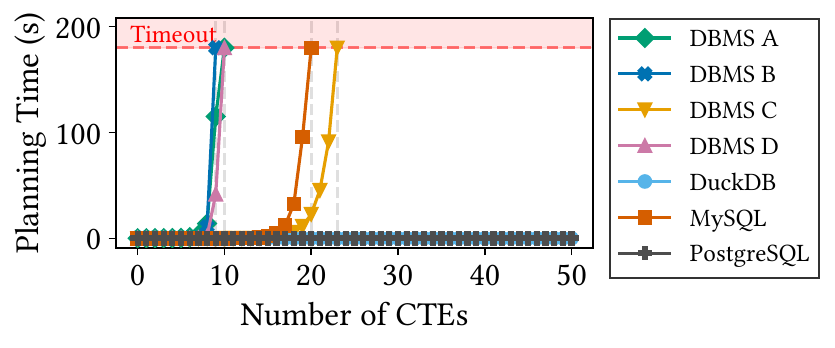}
  \vspace{-2.5em}
  \caption{All DBMSes except \duckdb{} and \postgres{} exhibit increases
  in planning times as the number of CTEs increases.}
  \label{fig:theme-cte-expansion}
\end{figure}

\Cref{fig:theme-cte-expansion} shows the query planning time as we increase the
number of CTEs to 50, using a three-minute planning timeout. \duckdb{} and
\postgres{} maintain low planning times, while the other five DBMSes all
eventually reach the planning timeout. \dbmsB{}, \dbmsD{}, and \dbmsA{}
reach timeouts around 10 CTEs whereas \mysql{} and \dbmsC{} reach timeouts
around 20 CTEs.

\begin{figure}[t]
  \vspace{-1em}
  \includegraphics[width=\columnwidth]{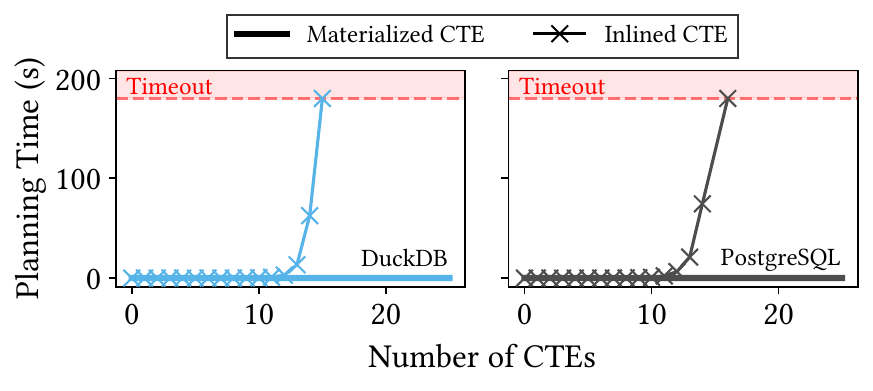}
  \vspace{-2.5em}
  \caption{\duckdb{} and \postgres{} exhibit increases in planning times when we
  force CTE inlining.}
  \label{fig:theme-cte-expansion-hint}
  \vspace{-0.5em}
\end{figure}

\sparagraph{Why does this pathology occur?}
We investigate this pathology by examining \duckdb{}'s and \postgres{}'s
planning behavior. \duckdb{} and \postgres{} avoid the planning-time blowup
because they materialize CTEs by default~\cite{duckdb-with,
postgres-materialized}. Under this strategy, the DBMS executes each CTE once,
stores the result in an intermediate table, and then references that table in
subsequent CTEs.
By using dialect-specific keywords, \Cref{fig:theme-cte-expansion-hint} compares
planning times when we force \duckdb{} and \postgres{} to materialize versus
inline the CTEs. Inlining causes planning time to increase exponentially because
each CTE references its predecessor twice, doubling the size of the query plan
representation at each level.

\subsection{Repeated Subquery Expressions}
Our next pathology involves repeated join subexpressions, as shown in
\Cref{lst:theme-join}. This query contains a subquery that joins the same table
to itself ten times. We repeat this subquery up to 20 times and measure the
planning time across all seven DBMSes. Importantly, each subquery contains only
a single eleven-table join graph, fewer tables than some queries in the Join
Order Benchmark~\cite{job}.

\begin{figure}[t]
  \includegraphics[width=\columnwidth]{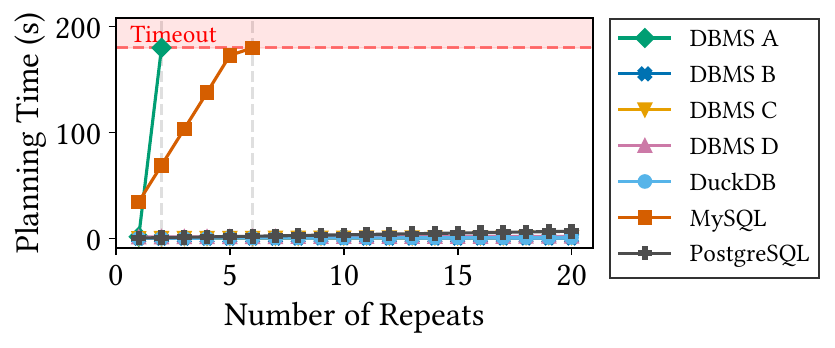}
  \vspace{-2.5em}
  \caption{\dbmsA{} and \mysql{} experience planning time increases as we
  repeat a join subquery.}
  \label{fig:theme-join}
  \vspace{-1em}
\end{figure}

\Cref{fig:theme-join} shows our results, again using a three-minute timeout.
\dbmsA{} and \mysql{} experience planning time increases as we increase the
number of repetitions, timing out at two and six repetitions respectively. The
other five DBMSes see modest increases in planning time, consistent with an
increasing query size.

\sparagraph{Why does this pathology occur?}
\mysql{}'s documented common-subexpression pass only targets boolean
expressions~\cite{mysql-cse}. We do not observe it re-using plans for these
repeated scalar subqueries, meaning it likely plans each repeated subquery
independently.
Like the other pathologies, \mysql{} performs exhaustive join order
enumeration by default; the single subquery instance is costly because it
contains eleven tables. These two factors explain both the noticeable planning
time with one repetition in \Cref{fig:theme-join} and the linear increase as we
add repetitions.
\dbmsA{} is closed-source, so we cannot inspect its code to investigate
why this pathology occurs.

\begin{listing}[t]
\begin{minted}{sql}
SELECT (SELECT COUNT(*) FROM lineitem l1, ..., lineitem l11
  WHERE l1.l_orderkey = l2.l_orderkey AND ...
    AND l10.l_orderkey = l11.l_orderkey) +
(SELECT COUNT(*) FROM lineitem l1, ..., lineitem l11
  WHERE l1.l_orderkey = l2.l_orderkey AND ...
    AND l10.l_orderkey = l11.l_orderkey) +
...
AS total
\end{minted}
\caption{The repeated join subexpression pathology. We vary the number of
repetitions from 1 to 20.}
\label{lst:theme-join}
\end{listing}

\begin{listing}[t]
\begin{minted}{sql}
SELECT COUNT(*)
FROM partsupp a, partsupp b, ..., partsupp n
WHERE (a.ps_partkey = b.ps_partkey
    OR a.ps_suppkey = b.ps_suppkey)
  AND (b.ps_partkey = ... OR b.ps_suppkey = ...)
  ...
  AND (n.ps_partkey = a.ps_partkey
    OR n.ps_suppkey = a.ps_suppkey)
  AND (a.ps_availqty > 100 OR c.ps_availqty > 100 OR ...)
\end{minted}
\caption{The disjunctive join pathology. We vary the size of the join
cycle up to 20 tables.}
\label{lst:theme-join-cycle}
\end{listing}

\subsection{Disjunctive Joins}
Another pathology we discovered involves joins whose conditions contain
disjunctions (i.e., use \texttt{OR}). \Cref{lst:theme-join-cycle} shows an
example: we join up to $N$ copies of the same table in a cycle, where the last
table joins to the first table, and add a final disjunctive predicate. We study
how the join size affects planning time by increasing the number of tables
involved up to 20.

\begin{figure}[t]
  \includegraphics[width=\columnwidth]{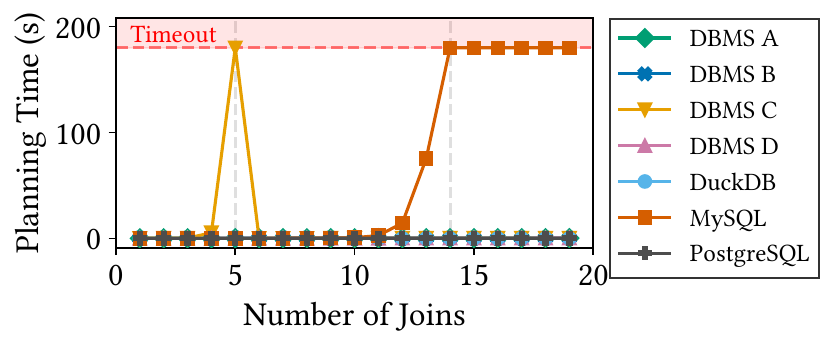}
  \vspace{-2.5em}
  \caption{\dbmsC{} and \mysql{} exhibit increases in planning times as we
  increase the size of the disjunctive join cycle. \dbmsC{} reverts to fast
  planning beyond five joins.}
  \label{fig:theme-join-cycle}
\end{figure}

\Cref{fig:theme-join-cycle} shows our results. Both \dbmsC{} and \mysql{}
exhibit planning time increases as the join cycle size increases. \dbmsC{}
times out on a query with 5 joins whereas \mysql{} times out at 14 joins.
Notably, \dbmsC{}'s planning time drops after five joins. We hypothesize this
drop occurs because an internal heuristic gives up on expensive planning beyond
a certain query complexity.

\sparagraph{Why does this pathology occur?}
We hypothesize that \dbmsC{} applies cost-based query rewrites on the
disjunctive predicates, which increases the number of plan candidates it
considers. For example, it could split a disjunction into two subqueries and
combine their results using a union. Because \dbmsC{} is closed-source, we
cannot inspect its code to identify the root cause.
\mysql{}'s planning time increases for the same reason as before: it performs
exhaustive join order enumeration by default. Consequently, its planning time
increases as the number of tables grows.

\begin{listing}[t]
\begin{minted}{sql}
-- May influence the physical plan (plan dependent)
SELECT COUNT(*) FROM part WHERE
  p_partkey = ASCII(MD5(REPEAT('a', 30000000))) + ...
  
-- Should not influence the physical plan (plan independent)
SELECT ASCII(MD5(REPEAT('a', 30000000))) + ...
\end{minted}
\caption{The constant folding pathology. Constant expressions may influence the
physical plan or be independent.}
\label{lst:theme-const-fold}
\end{listing}

\begin{figure}[t]
  \vspace{-1em}
  \includegraphics[width=\columnwidth]{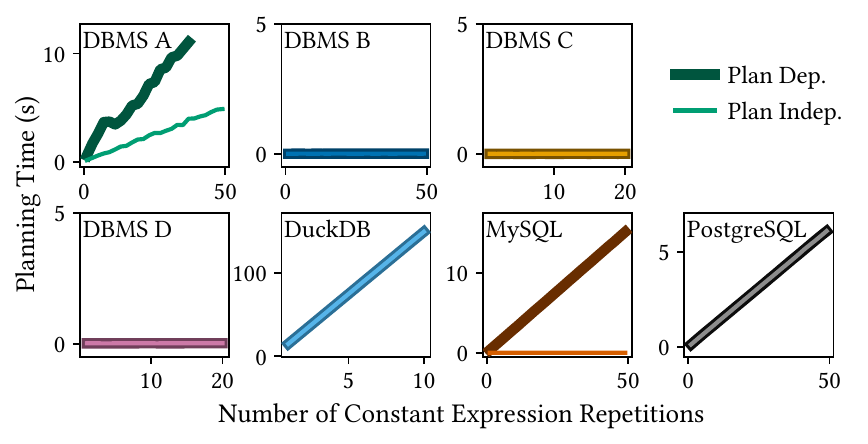}
  \vspace{-2.5em}
  \caption{Planning time as we vary the number of constant expressions
  in plan dependent and independent locations.}
  \label{fig:theme-const-fold}
  \vspace{-1em}
\end{figure}

\subsection{Adversarial Constant Folding}
Finally, we examine constant folding, an example of which is in
\Cref{lst:theme-const-fold}. The slow-to-plan \duckdb{} and
\postgres{} queries in \Cref{fig:cross-plot} contain expensive constant
expressions that both systems appear to evaluate during planning.

To study this behavior, we repeatedly add an expensive constant expression to
either the \texttt{SELECT} list (``plan independent'') or \texttt{WHERE} clause
(``plan dependent''). \Cref{fig:theme-const-fold} shows our results. Because the
DBMSes support different functions, we use a different expression on each
system. Thus the absolute planning times and repetition counts are not directly
comparable; we instead focus on how planning time changes within each DBMS.

Planning time grows with the repetition count on \dbmsA{}, \duckdb{},
\mysql{}, and \postgres{}, suggesting that these systems evaluate expressions
during planning. \mysql{} appears to distinguish expressions that affect plan
choice from those that do not: planning time grows when expressions appear
in the \texttt{WHERE} clause, but remains stable when they appear in the
\texttt{SELECT} list. \dbmsB{}, \dbmsC{}, and \dbmsD{} show no planning time
increases, suggesting that they do not execute these expressions during
planning.

Constant folding during planning can improve query execution when an expression
affects plan choice or its result can be reused. However, our results show that
indiscriminately folding every constant expression can also stall the planner.
Planners should consider the optimization benefit against the cost of folding
during planning, particularly if the query may not be executed immediately.

%% file: sections/4-related.tex
\section{Related Work}\label{sec:related}
\sparagraph{Query planner performance.}
OptMark~\cite{optmark-li16} is a benchmark toolkit that evaluates query planning
effectiveness and efficiency. It measures efficiency using logical counts, e.g.,
the number of join orders explored (among others), which correlates with
optimization time. While OptMark is a methodology for comparing planners on
efficiency, our work focuses on finding and analyzing pathological queries that
cause long planning times.
Ilyas et al. explore predicting query planning time, which might help avoid
pathological queries~\cite{planest-ilyas03}.
Orthogonal work also proposes techniques to improve the performance of query
planning~\cite{parqo-han08, parqo2-trummer16, adaptiveopt-neumann18, bqo-fent23,
simplicity-hertzschuch21}.

\sparagraph{Database performance bug finding.}
Prior works propose techniques for finding query execution performance bugs in
database systems~\cite{apollo, amoeba, cert, puppy, hulk, short-circuit}.
The high-level idea is to identify a general property about a pair of queries
(e.g., making a query more restrictive should result in an output cardinality
estimate no larger than the original query~\cite{cert}). Then, they generate
query pairs and check if they satisfy the property and report a bug if not.
Our work differs, as we identify and study queries that trigger query planner
performance bugs (instead of query execution performance bugs).

\sparagraph{Database correctness bug finding.}
A complementary set of papers propose techniques to find correctness bugs in
database systems~\cite{argus, query-plan-guidance, differential-query-plans,
pivoted-query-synth, nonopt-ref-construction, query-partitioning, const-opt},
applying a similar approach grounded in query pairs as in performance bug finding.
Among these proposals, Argus is most thematically similar to this work as it
uses LLMs to generate pairs of semantically equivalent but syntactically
different queries that they then execute and compare results~\cite{argus}.
Our work differs as we are searching for and analyzing query planner performance
bugs, not correctness bugs.

\sparagraph{Agentic performance debugging.}
Researchers have recently been exploring leveraging LLMs to help with
performance debugging~\cite{panda-debug} and diagnostics~\cite{dbot-diagnosis}.
Such techniques propose workflows involving multiple agents, grounded in
artifacts (e.g., prior customer tickets, troubleshooting
documents)~\cite{panda-debug} and equipped with tools~\cite{dbot-diagnosis}, to
propose remediations for a database performance problem~\cite{panda-debug} or
diagnostics reports~\cite{dbot-diagnosis}.
Our work differs as we leverage agents to search for queries that trigger query
planning stalls, but we are not using the agents to propose fixes.

%% file: sections/5-conclusion.tex
\section{Conclusion}\label{sec:conclusion}
This paper showed that query planning is not always a low-cost operation. Using
a lightweight agentic search, we found queries that take more than three minutes
to plan on each of the seven DBMSes.
Most slow-to-plan queries are DBMS-specific, possibly because they expose
planner-specific implementation quirks. Nevertheless, our analysis showed that
some query pathologies affect multiple DBMSes.
Our findings have implications for planner implementers and downstream systems.
Planner implementers should account for adversarial inputs and design their
planners to have stable planning performance. Systems that invoke query planners
or depend on physical query plans should not assume that planning is always
inexpensive and should provide a fast fallback when planning exceeds a time
budget.
More broadly, we hope these findings encourage further study of query planning
latency as a first-class performance and robustness concern.